\documentclass{svproc}
\usepackage{url}
 \usepackage{amsmath}
 \usepackage{lineno}
 \usepackage{overpic,color}
 \usepackage{graphicx}

\begin{document}
\mainmatter              
\title{Probing QCD matter via $K^{*0}(892)$ and $\phi(1020)$ resonance production at RHIC}
\titlerunning{$K^{*0}(892)$ and $\phi(1020)$ resonance production}  
%
\author{Md Nasim (for the STAR Collaboration) }
\authorrunning{Md Nasim} 
%
%
\institute{Indian Institute of Science Education and Research Berhampur, \\Odisha 760010, India.\\
\email{nasim@iiserbpr.ac.in}}

\maketitle              


\begin{abstract}

We present the measurements of invariant yields of $K^{*0}$ and $\phi$ resonances  at midrapidity in Au+Au collisions at $\sqrt{s_{NN}}$ = 7.7 - 200 GeV using the STAR detector. 
The transverse momentum ($p_{T}$) spectra and $p_{T}$-integrated yields of $K^{*0}$ and $\phi$ have been studied. The ratios between resonance ($K^{*0}$ and $\phi$) to non-resonance particles ($K$) are presented as a function of centrality. It is found that $K^{*0}/K^{-}$ ratios are suppressed in the most central collisions as compared to peripheral ones for all studied collision energies. On the other hand, $\phi/K^{-}$ ratios are weakly depend on centrality. These results can be understood  by considering  the effect of more hadronic rescattering for  $K^{*0}$ (lifetime $\sim$ 4 fm/c) as compared to $\phi$ (lifetime $\sim$ 42 fm/c).  
We have also presented  the measurement of the first-order azimuthal anisotropy (known as directed flow, $v_{1}$) of $\phi$ meson as a function of rapidity ($y$) at midrapidity in Au+Au collisions at $\sqrt{s_{NN}}$ = 7.7 - 200 GeV. 
The slope of $\phi$-meson $v_{1}$ ($dv_{1}/dy$) has been compared to the $dv_{1}/dy$ of other identified particles. We have found that all particles that consist of produced quarks show similar behaviour for $\sqrt{s_{NN}}$ $>$ 14.5 GeV.

\end{abstract}
\section{Introduction}
The aim of the STAR experiment at Relativistic Heavy Ion Collider (RHIC) is to study the QCD matter by colliding nuclei at ultra-relativistic speeds~\cite{starwhite}. 
The study of $K^{*0}$ (lifetime $\sim$ 4 fm/c)  and $\phi$ (lifetime $\sim$ 42 fm/c) production in heavy-ion collisions can be used to probe the medium created after the collision~\cite{kstar_star}. 
The $K^{*0}$ resonance has a short  lifetime, therefore it may decay during the hadronic phase and its decay products may undergo elastic or pseudoelastic scatterings~\cite{pseudo}. Due to elastic or pseudoelastic scatterings, the final yield of  $K^{*0}$ resonance may get changed.  The $K^{*0}$ resonance yields may get reduced due to rescattering of its daughters through elastic scattering or the yields may be regenerated through pseudoelastic scattering between chemical and kinetic freeze-out~\cite{kstar_star}. However, due to longer lifetime as compared to the fireball, $\phi$ mesons mostly decay outside of the fireball and its daughters are not affected by late-stage hadronic scatterings. Hence, the study of $K^{*0}$ and $\phi$ resonances provides an important information about the late-stage hadronic scatterings. \\
In high-energy heavy-ion collisions, particles are produced with an azimuthally anisotropic momentum distribution.  Directed flow ($v_{1}$) is a measure of azimuthal angular anisotropy of the produced particles with respect to the first-order event plane~\cite{flow}.  The $v_{1}$ is an initial state effect and expected to be sensitive to the equation of state of the system formed in the collision.  The $\phi$ meson freezes out early and is expected to have small hadronic interaction cross section. The measured $v_{1}$ of $\phi$ meson can be used as a clean probe to study the QCD matter~\cite{smallx}.
 
\section{Data sets and methods}
\begin{figure}
\center{
\begin{overpic}[scale=0.42]{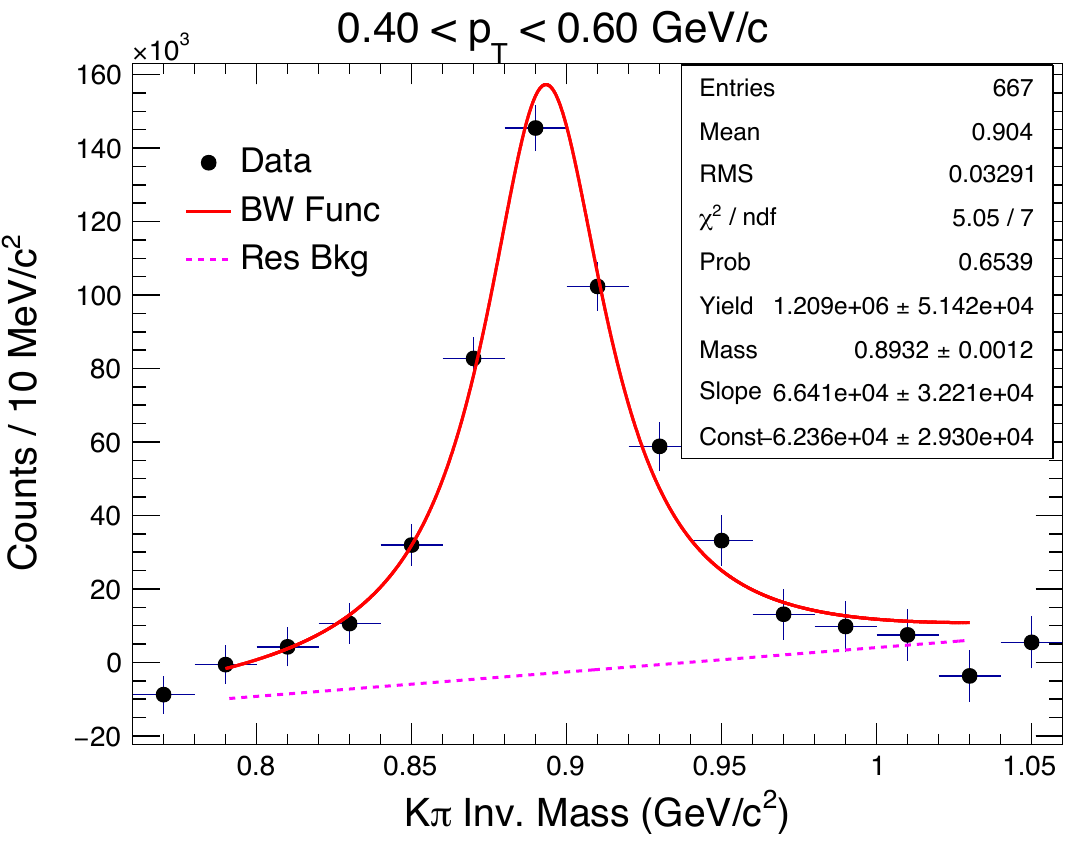}
\end{overpic}
\begin{overpic}[scale=0.4]{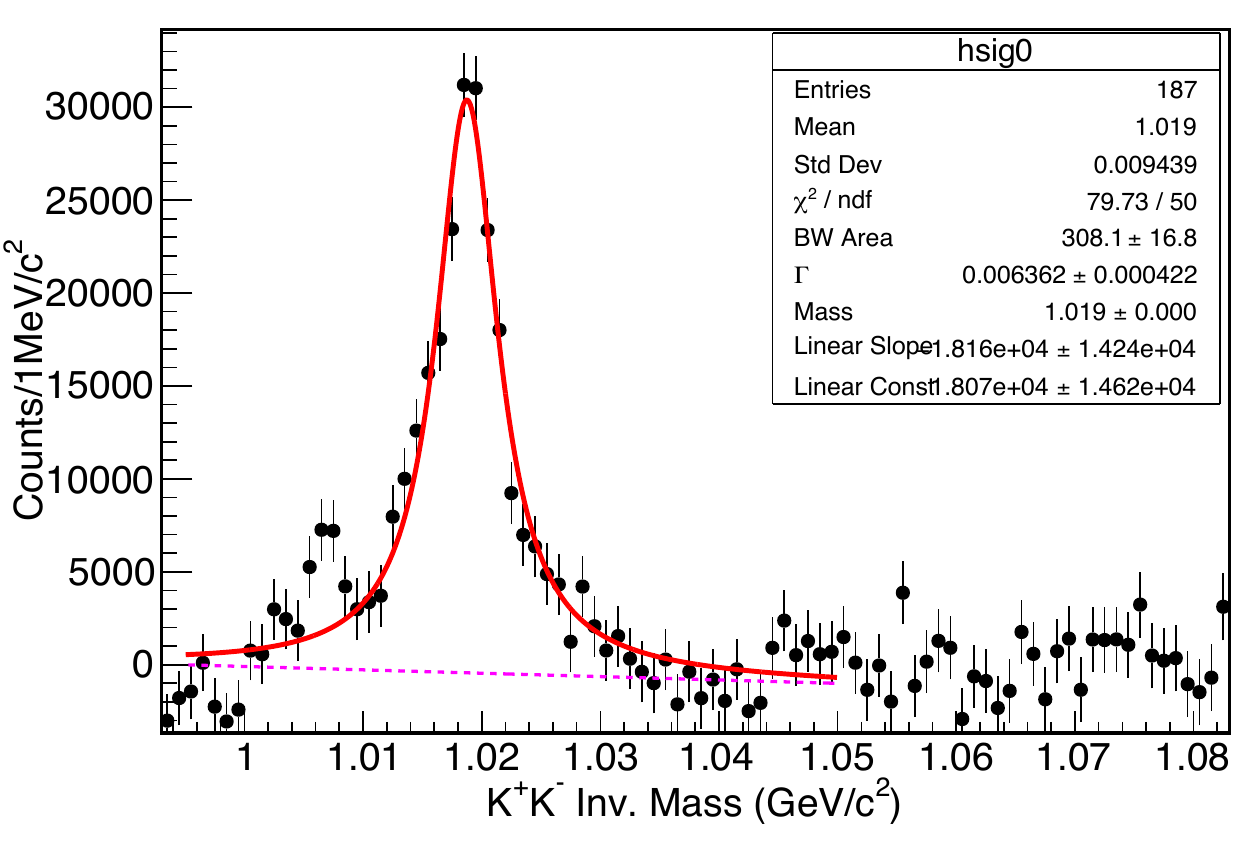}
\put (20,70) {\small STAR Preliminary}
\end{overpic}
}

        \caption{\label{fig:1} Invariant mass distribution of $K^{\pm}\pi^{\mp}$  and $K^{+}K^{-}$ pairs after mixed event background subtraction for miminum bias (0-80\%) Au+Au collisions at  $\sqrt{s_{NN}}$ =11.5 GeV for  0.4 $<$ $p_{T}$ $<$$0.6$ GeV/c.}
      \end{figure}

The results presented here are based on data collected at $\sqrt{s_{NN}}$ = 7.7, 11.5, 14.5, 19.6, 27, 39, 62.4 and 200 GeV  in Au+Au collisions by the STAR detector using  a minimum-bias trigger. 
The Time Projection Chamber (TPC)~\cite{tpc} and Time of Flight (TOF)~\cite{tof} detectors with full 2$\pi$ coverage are used for particle identification in the central pseudorapidity ($\eta$) region ($|\eta|$$<$ 1.0). The
$K^{*0}$ and $\phi$  resonances  are reconstructed from the following  hadronic decay channels: $K^{*0}$ $\longrightarrow$  $K^{\pm}$ +$\pi^{\mp}$ and $\phi$ $\longrightarrow$ $K^{+}$ + $K^{-}$. Event mixing technique has been used for combinatorial background estimation~\cite{besphi}. Figure~\ref{fig:1} shows invariant mass distribution of $K^{\pm}\pi^{\mp}$  and $K^{+}K^{-}$ pairs after mixed event background subtraction.  \\
The first harmonic coefficient of the Fourier decomposition of azimuthal distribution with respect to the first-order event plane angle ($\psi_{1}$) can be expressed as $v_{1} = \langle \rm{cos}(\varphi-\psi_{1}) \rangle$, 
where $\varphi$ is the azimuthal angle of the produced particle. The first-order event plane angles are calculated using a forward rapidity detectors (Beam-Beam Counters or Zero-Degree Calorimeters)~\cite{zdc}. The measured $v_{1}$ with respect to the first-order event plane has been corrected for the finite event plane resolution. The more details about $v_{1}$ measurement in STAR can be found in Ref.~\cite {besv1}. 

\section{Results}
\begin{figure}
 \center{\begin{overpic}[scale=0.46]{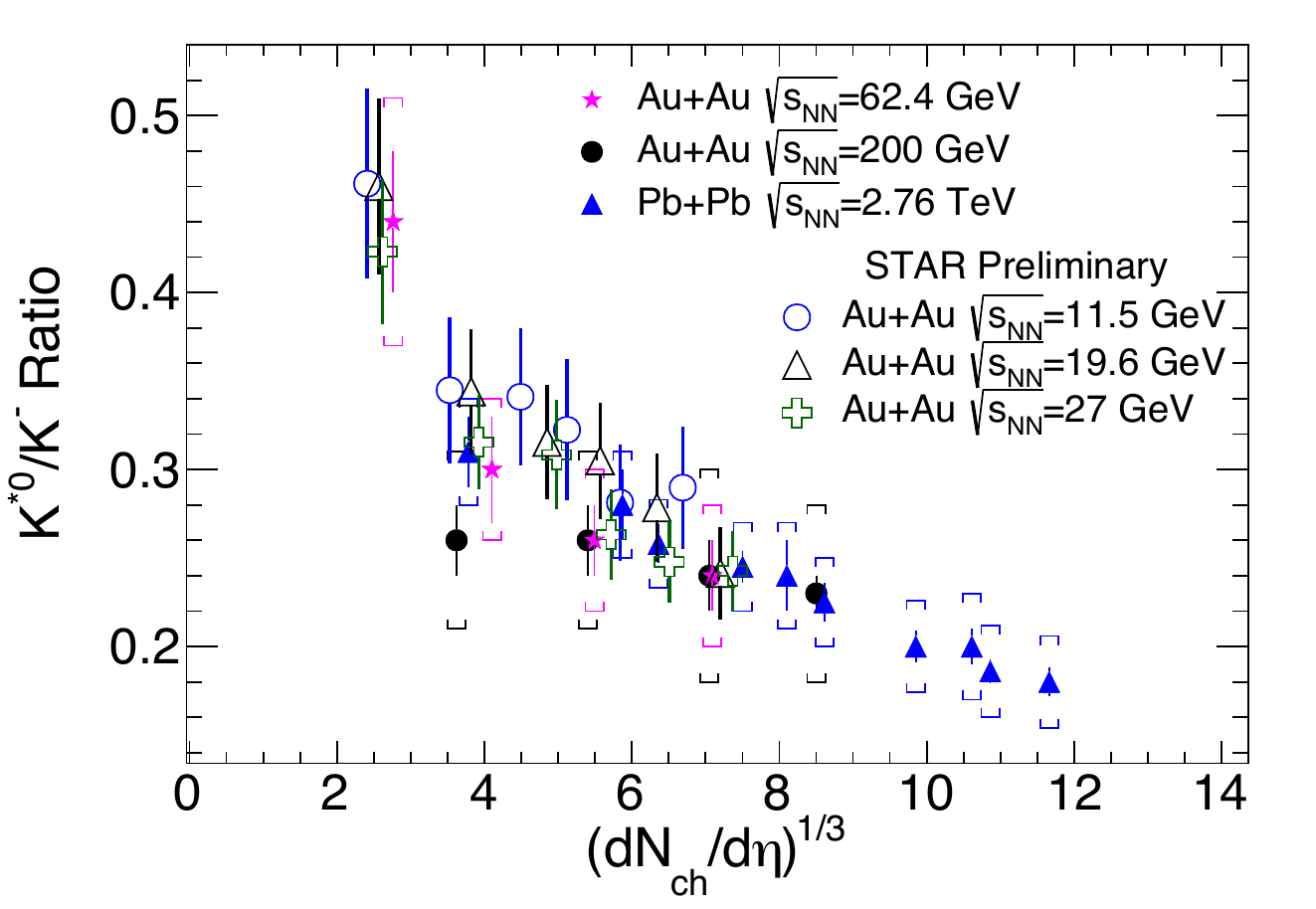}
\end{overpic}
\begin{overpic}[scale=0.28]{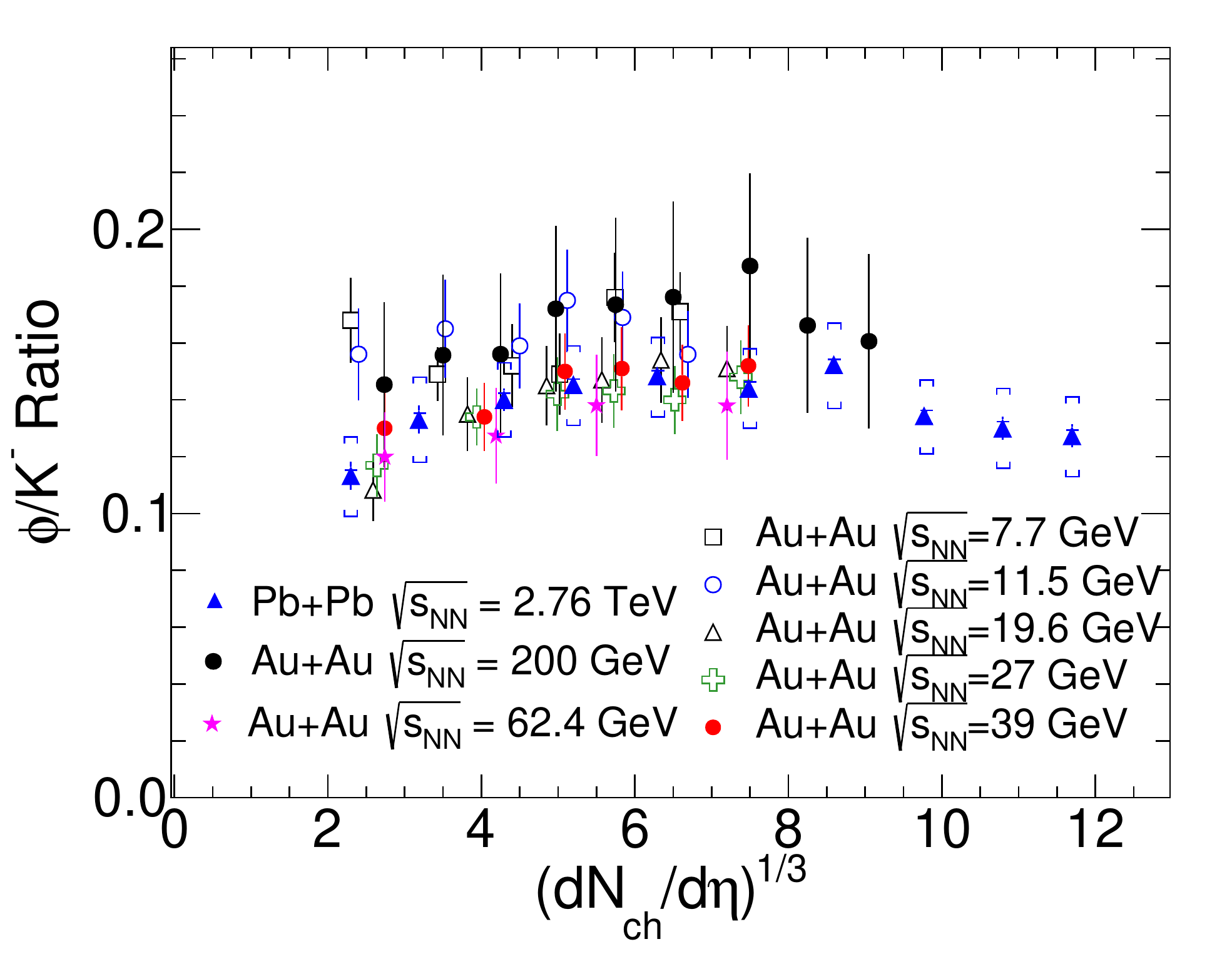}
\put (20,65) {\tiny STAR Preliminary}
\end{overpic}
}

        \caption{\label{fig:2} Ratios $K^{*0}/K^{-}$  and $\phi/K^{-}$ as a function of $(dN_{ch}/d\eta)^{1/3}$  in Au+Au and Pb+Pb collisions at various  $\sqrt{s_{NN}}$~\cite{besphi,bespikp,62gevkstar,200gevphi,2760gevphi}.  }
      \end{figure}

Figure~\ref{fig:2} shows $K^{*0}/K^{-}$  and $\phi/K^{-}$ ratios as a function of $(dN_{ch}/d\eta)^{1/3}$ (the cubic root of the charged-particle multiplicity density measured at midrapidity)  in Au+Au collisions at various center-of-mass energies measured by STAR experiment~\cite{besphi,bespikp,62gevkstar,200gevphi,2760gevphi}. The results are compared to the measurements performed at $\sqrt{s_{NN}}$ =2.76 TeV in Pb+Pb collision by the ALICE collaboration. The $K^{*0}/K^{-}$  ratios decrease with increasing $(dN_{ch}/d\eta)^{1/3}$ (proxy for system size) for all studied collisions energies.  On the other-hand, $\phi/K^{-}$ ratios are almost independent of $(dN_{ch}/d\eta)^{1/3}$. The observed suppression of the $K^{*0}/K^{-}$  ratios  could be due to rescattering effect as discussed earlier.

\begin{figure}
        \center{\includegraphics[scale=0.3]{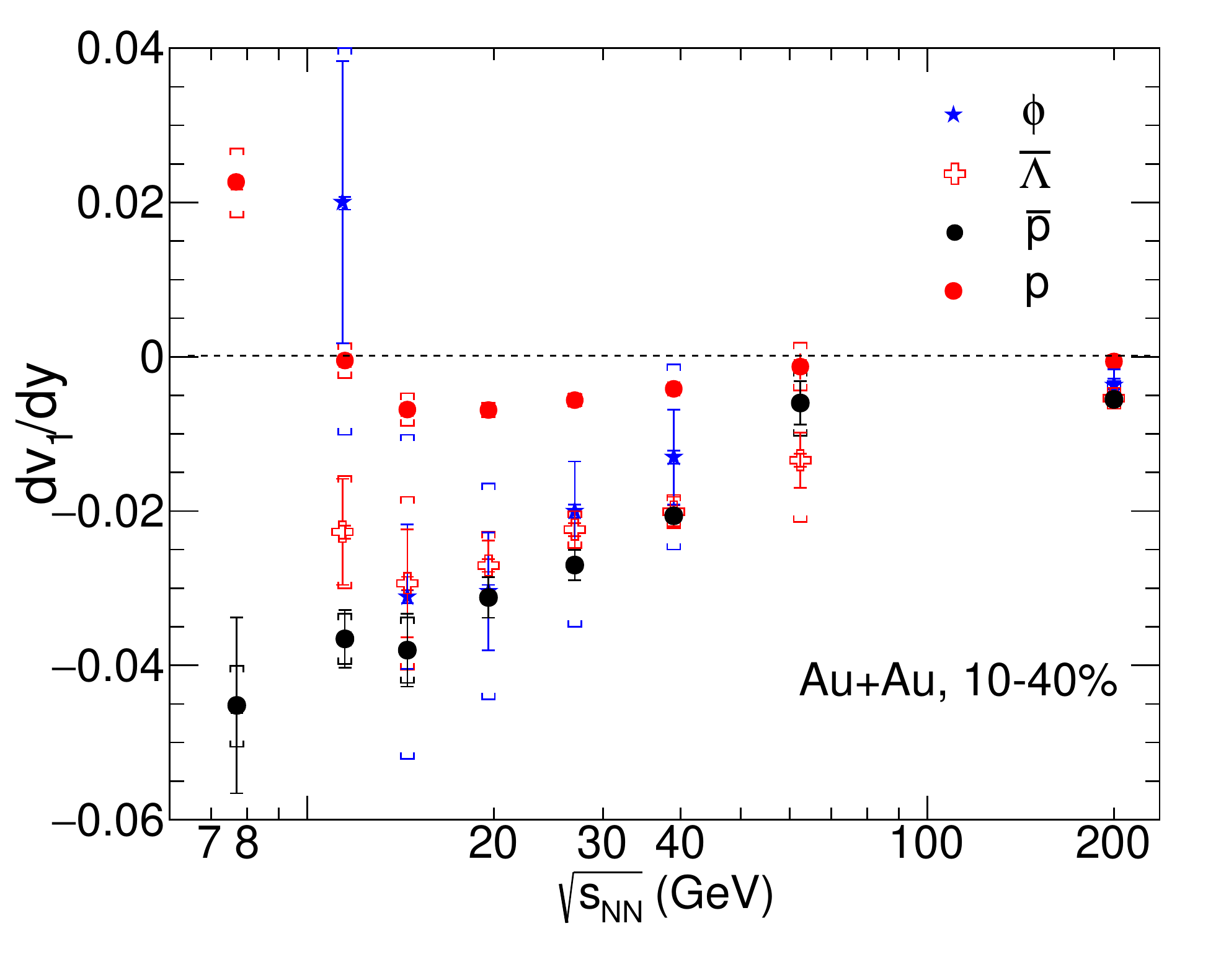}
        }
        \caption{\label{fig:3} Directed flow slope ($dv_{1}/dy$) versus beam energy for 10-40\% centrality in Au+Au collisions for $\phi$, $\bar{\Lambda}$ , $p$ and $\bar{p}$~\cite{besv1}.}
      \end{figure}

Directed flow slope, $dv_{1}/dy$, versus beam energy for  $\phi$, $\bar{\Lambda}$ , $p$ and $\bar{p}$  is presented in Fig.~\ref{fig:3}  for 10-40\% central Au+Au collisions~\cite {besv1}.
The slope for  $\bar{\Lambda}$  is negative for all energies and is consistent within errors with that for $\bar{p}$. The $\bar{p}$, $\phi$ and $\bar{\Lambda}$ are seen to have similar $v_{1}(y)$ for $\sqrt{s_{NN}}$ $>$ 14.5 GeV. All these species consist from quarks that are produced in the collision.
For $\sqrt{s_{NN}}$ $<$ 14.5 GeV, the current statistical uncertainty is too large to make any conclusion. 

\section{Summary}
We report the measurement of $K^{*0}$ and $\phi$ resonance production  at midrapidity in Au+Au collisions at  $\sqrt{s_{NN}}$ = 7.7-200 GeV recorded by the STAR detector. 
The $K^{*0}/K^{-}$  and $\phi/K^{-}$ ratios as a function of $(dN_{ch}/d\eta)^{1/3}$ is presented for different center-of-mass energies. The $K^{*0}/K^{-}$ ratios are found to decrease with increasing $(dN_{ch}/d\eta)^{1/3}$. 
Whereas, $\phi/K^{-}$ ratios are nearly independent of  $(dN_{ch}/d\eta)^{1/3}$. The observed suppression of  $K^{*0}/K^{-}$ ratios in central collisions could be due to the effect of hadronic rescattering which reduce the measured yield of short-lived resonances. The directed flow of $\phi$ meson is presented  for different collision energies. Directed flow slope ($dv_{1}/dy$)  of $\phi$ meson is found to be consistent within errors with that of  $\bar{\Lambda}$ and $\bar{p}$ for $\sqrt{s_{NN}}$ $>$ 14.5 GeV.

%

\end{document}